\documentclass{jpconf}
\usepackage{amssymb}
\usepackage{graphicx}

\def\gagg{g_{a\gamma}}
\begin{document}
\title{Status report of the Tokyo axion helioscope experiment}
\author{Y Inoue$^1$,
  M Minowa$^2$, Y Akimoto$^2$, R Ota$^2$, T Mizumoto$^2$
  and~A~Yamamoto$^3$}
\address{$^1$ International Center for Elementary Particle Physics,
  University of Tokyo,
  7-3-1 Hongo, Bunkyo-ku, Tokyo 113-0033, Japan}
\address{$^2$ Department of Physics, School of Science, University of Tokyo,
  7-3-1 Hongo, Bunkyo-ku, Tokyo 113-0033, Japan}
\address{$^3$ High Energy Accelerator Research Organization (KEK),
  1-1 Oho, Tsukuba, Ibaraki 305-0801, Japan}
\ead{berota@icepp.s.u-tokyo.ac.jp}
\begin{abstract}
We have searched for solar axions with a detector
which consists of a $4\,{\rm T}\times2.3\,{\rm m}$
superconducting magnet, PIN-photodiode X-ray detectors,
and an altazimuth mount to track the sun.
The conversion region is filled with cold helium gas which modifies the
axion mass at which coherent conversion occurs.
In the past measurements, axion mass from 0 to 0.27eV have been scanned.
Since no positive evidence was seen, an upper limit to the
axion-photon coupling constant was set to be
$\gagg<6\hbox{--}10\times10^{-10}\rm/GeV$
(95\% CL) depending on the axion masses.
We are now actively preparing for a new
stage of the experiment aiming at one to a few eV solar axions.
In this mass region, our detector might be able to
check parameter regions which are preferable to the
axion models.
\end{abstract}
\section{Introduction}
\begin{figure}
  \begin{minipage}[b]{0.45\textwidth}
    \includegraphics[width=\textwidth]{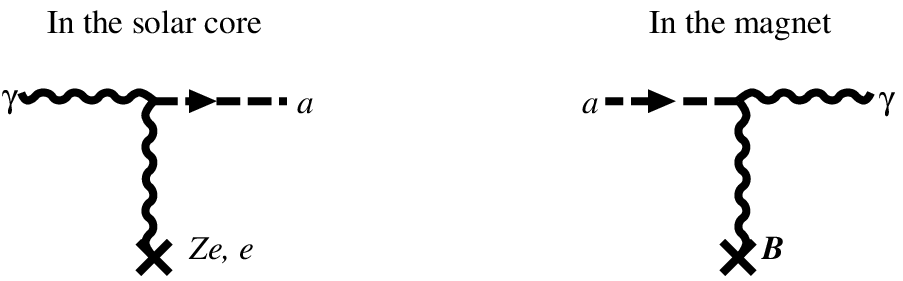}
    \caption{The solar axions produced via the Primakoff process
      in the solar core are, then, converted into X-rays
      via the reverse process in the magnet.}
    \label{fig:principle}
  \end{minipage}
  \hskip0.1\textwidth plus 1fil minus 5pt
  \begin{minipage}[b]{0.45\textwidth}
    \includegraphics[width=\textwidth]{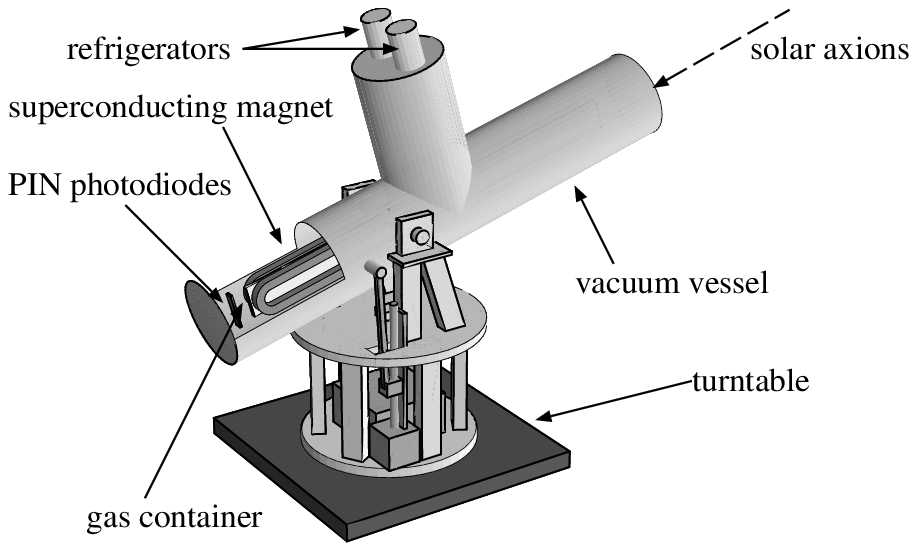}
    \caption{The schematic view of the axion helioscope.}
    \label{fig:sumico}
  \end{minipage}
\end{figure}

Axions \cite{pq1977} that are thermally produced
in the core of the sun through the Primakoff process
(Fig. \ref{fig:principle})
are called the solar axions.
Sikivie \cite{sikivie1983} proposed an ingenious experiment to detect
such axions.
The detection device called axion helioscope
is a system of a strong magnet and an X-ray detector,
where the solar axions is converted into X-ray photons
through the inverse Primakoff process (Fig. \ref{fig:principle})
in the magnetic field.
Conversion is coherently enhanced even for massive axions
by filling the conversion region with light gas.
If the axion mass $m_a$ is at around a few eV, 
detection of the solar axions becomes feasible.

We have constructed
an axion helioscope with a dedicated superconducting magnet.
Cold helium gas was used as the conversion medium,
and the highest axion mass was targeted at 2.6\,eV.
We report the current status of this experiment.
\section{The Sumico V detector}
The schematic figure of the axion helioscope is illustrated
in Fig.~\ref{fig:sumico}.
It consists of
a superconducting magnet, X-ray detectors, a gas container,
and an altazimuth mounting.

The magnet \cite{sato1997}
consists of two 2.3-m long
race-track shaped superconducting coils running parallel
with a 20-mm wide gap between them.
The transverse magnetic field in the gap is 4\,T.
The magnetic field can be maintained without an external power supply
with a help of a persistent current switch.
The magnet is kept lower than 6\,K by two Gifford-McMahon refrigerators.
The container to hold dispersion-matching gas is inserted
in the aperture of the magnet.
It is made of four 2.3-m long stainless-steel square pipes 
and 5N high purity aluminium sheets wrapping around them
to achieve high uniformity of temperature.
The measured thermal conductance between the both ends was
$1\times10^{-2}\mathrm{W/K}$ at 6\,K under 4\,T.
The ``solar'' end of the gas container
is suspended by three Kevlar cords.
%The heat injection through the cords is negligible.
The ``detector'' end at the opposite side
is flanged to be fixed to the magnet.
This end is terminated
with an X-ray window which is transparent above 2\,keV
and can hold gas up to 0.3\,MPa.
The gas introducing pipelines are also at this side.
The converted X-ray is viewed by
sixteen PIN photodiodes.
Details on the X-ray detector are given in Ref.\ \cite{naniwaPIN}.
They are constructed in a vacuum vessel
which is mounted on an altazimuth mount
to track the sun.
It can track the sun about a half of a day.
During the other half of a day, background spectrum is measured.

\section{Past and future measurements}
\subsection{Phase I --- the first measurement}
Phase I of the solar observation was performed
from 26th till 31st December 1997
without the gas container  \cite{sumico1997}.
Since the conversion region was vacuum,
the sensitivity was limited below $m_a<0.03\rm\,eV$.
From the absence of an axion signal,
an upper limit on the axion-photon coupling is given to be
$\gagg<6.0\times10^{-10}\rm GeV^{-1}$ (95\% CL).
\subsection{Phase II --- with low density helium gas}
\begin{figure}[t]
  \begin{minipage}[t]{0.43\textwidth}
    \hrule height 0pt
    \includegraphics[width=\textwidth]{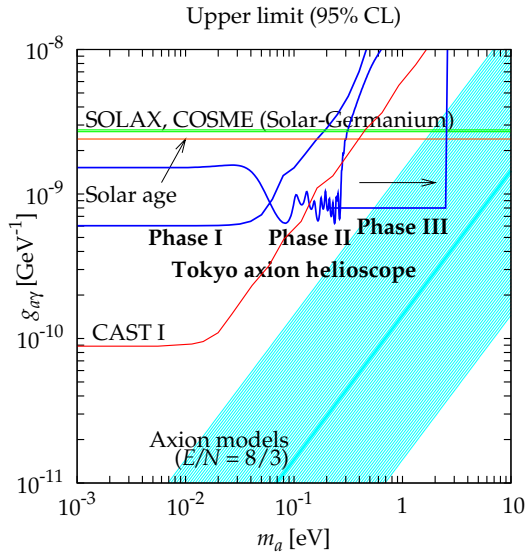}
  \end{minipage}
  \hskip 0.1\textwidth plus 1fil minus 4pt
  \begin{minipage}[t]{0.47\textwidth}
    \hrule height 0pt
    \caption{Exclusion limit on $\gagg$ versus $m_a$
      at 95\% confidence level.
      Phase I and II denote the past results from this experiment,
      and Phase III shows the area where is aimed at the next stage.
      The limits inferred from the solar age consideration,
      the limits from other experiments:
      SOLAX \cite{solax1999},
      COSME \cite{cosme2002}
      and CAST \cite{cast2007} are show together.
      The shaded area shows the preferred axion models.
    }
    \label{fig:limit}
  \end{minipage}
\end{figure}

Phase II was performed from July to September 2000,
where low density helium was introduced \cite{sumico2000}.
The sensitive mass region was brought up by an order of magnitude.
Again, no signature of axions was seen,
restricting the axion-photon coupling to be
$\gagg<(\hbox{6.2--10.4})\times10^{-10}\rm GeV^{-1}$ for
$0.05<m_a<0.27\rm\,eV$.
The exclusion limit at 95\% confidence level obtained
from Phase I and II are
shown together in Fig.\ \ref{fig:limit}.
\subsection{Phase III --- the next stage}
At present, we are preparing for the next stage.
In Phase III,
high density helium will be introduced
aiming at more massive axions
up to $m_a\lesssim\hbox{2.2--2.6\,eV}$.
To achieve this goal, several developments have been made.
When the superconducting magnet quenches, the high-density helium should
be relieved safely before the gas container would burst.
The gas introducing pipeline was completely reworked based on the measured
pressure change after a forced quench with a safely low pressure helium.
In addition, a cryogenic rupture disk was installed as a last resort.
Another challenge is that
the higher the axion mass becomes,
the narrower the resonance width of the conversion becomes.
The gas density should be controlled
with an extreme accuracy
typically of the order of 0.1\% at $m_a=2.6\rm\,eV$,
and many data points should be measured to scan a range.
For this purpose, an automated gas controlling system was developed.
It has successfully demonstrated to stabilize the temperature
and to control the pressure of helium gas
with a good accuracy.
\section{Conclusions}
We have searched for solar axions
with an axion helioscope,
and axion mass from 0 to 0.27eV have been scanned.
No evidence of axion signal has been seen so far,
implying an exclusion limit
to the axion-photon coupling
as shown in Fig.\ \ref{fig:limit}.
In the lower mass region, the above limit has been renewed by CAST.
However, we are preparing for the next stage
aiming at a higher mass region
where our detector might be able to 
check the parameter regions preferred by theory.
In December 2007, a preliminary measurement started at $m_a\simeq1\rm\,eV$.
Remaing problems are being resolved one by one.
The ultimate density will be reached in 2008, although a full range scan
will take years.
\ack
The authors thank the ex-director general of KEK, Professor H. Sugawara,
for his support in the beginning of the experiment.
The authors are also thankful to
the support from the Grant-in-Aid for COE research
by the Japanese Ministry of Education, Science, Sports and Culture,
and the support from the Matsuo Foundation.
\section*{References}
\bibliography{proc}
\end{document}